\documentclass[pra, showpacs, twocolumn]{revtex4}
\usepackage{amssymb}
\usepackage{bm}
\begin{document}
\title
{Diagnostics of macroscopic quantum states of Bose-Einstein
condensate in double-well potential by nonstationary Josephson
effect}
\author{E.\,D.\,Vol}
\email{vol@ilt.kharkov.ua}
\affiliation{%
B.\,I.\, Verkin Institute for Low Temperature Physics and
Engineering National Academy of Sciences of Ukraine, Lenin av. 47
Kharkov 61103, Ukraine}

\date{\today}

\begin{abstract}
We propose a method of diagnostic of a degenerate ground state of
Bose condensate in
 a double well potential. The method is based on the study of
 the one-particle coherent tunneling under switching the
 time-dependent weak Josephson coupling between the wells. We
 obtain a simple expression that allows to determine the phase of
 the condensate and the total number of the particles in the
 condensate from the relative number of the
 particles in two wells $\Delta n =n_1-n_2$ measured before the
 Josephson coupling is switched on and after it is switched off.
 The specifics of the application of the method in the cases of
 the external and the internal Josephson effect are discussed.
\end{abstract}


\maketitle

Beginning from its first observation \cite{1} the Bose-Einstein
condensation (BEC) of atoms in alkali metal vapours remains the
source for new possibilities for the study of macroscopic quantum
phenomena. One of these phenomena is the coherent tunneling of
atoms between two coupled Bose condensates (BC)  \cite{2}, that is
analogous to the Josephson effect in superconductors. It is known
\cite{3} that for the case when the total number of atoms
 in the trap $N=n_1+n_2$ is conserved and the trap is symmetric relative
two BC the average relative number of atoms
$\overline{n_1-n_2}=\langle\Psi|\hat{n}_1-\hat{n}_2|\Psi\rangle$
is equal to zero in the ground state and in any excited state.
Therefore, one can expect that in such a situation the study of
nonstationary coherent tunneling (which is realized when one or
several parameters of the system depends on time) is more
informative for the diagnostics of the macroscopic wave functions
of the condensates. In this case the average value of the relative
number of the atoms $\overline{n_1-n_2}$ measured in the time
$t_0$ is generally nonzero one and it depends on the history of
the systems at all $t<t_0$. In this paper we show that
nonstationary Josephson effect can be used for the diagnostics of
a macroscopic state of BC and the total number of the atoms in the
condensate.

We consider the simple model of coherent tunneling between two BC,
described in \cite{4} (see also references in it). The model is
based on the two mode approximation, that implies that each of $N$
bosons can be in one of two states and the dynamical coupling
between these state allows the bosons to jump from one state to
the other. Such a model is applicable for a description of the
external as well as the internal Josephson effect in Bose systems.
The external Josephson effect \cite{5} can be realized if the Bose
gas is confined in a double-well trap and the tunneling between
two wells is small. In this case two modes correspond to self
consistent ground states in each well. The internal Josephson
effect \cite{6} can be realized in a Bose gas with two
macroscopically occupied hyperfine states (e.g. the $|F=1,
m_F=-1\rangle$ and $|F=2, m_F=1\rangle$ states of $^{87}$Rb
atoms). The dynamical coupling between the two states is settled
by a resonant laser field applied to the system. At the beginning
we specify the simplest case of the external Josephson effect at
$T=0$.

The Hamiltonian of the symmetric two-mode model has the form
\begin{equation}\label{1}
  H=H_0^s+H_1(t)\equiv\frac{K}{8}(\hat{n}_1-\hat{n}_2)^2-\frac{E_J(t)}{2}(a_1^+a_2
  +h.c),
\end{equation}
where $a_i^+$ ($a_i$) are the  creation (annihilation) operators
for the well $i$, and $\hat{n}_i=a_i^+a_i$ are the number
operators. The parameter $K$ is determined by the interaction
between the atoms in the well. Here we consider the case of $K>0$
that corresponds to the repulsive interaction. The value of $E_J$
is determined by the overlapping of the wave functions of two
modes  and it can be controlled by a variation of the height and
(or) width of the barrier. For the external Josephson effect
without loss of generality one can set the Josephson coupling
$E_J(t)$ be real.

Let us consider the situation when the dynamical coupling between
two condensates is switched on at $t=t_i=0$ and switched off at
$t=t_f$. At $t<0$ and $t>t_f$ the coupling parameter $E_J(t)=0$
and the occupation numbers operators $\hat{n}_1$ and $\hat{n}_2$
as well as the relative number operator $\hat{n}_1-\hat{n}_2$
commute with the Hamiltonian and do not depend on time. During the
time when the coupling is switched on the operator
$\hat{n}_1-\hat{n}_2$ is changed. Let at $t=0$ the wave function
of the two mode Bose condensate is $\Psi(0)$ and at $t=t_f$ it
becomes $\Psi(t_f)$. The task we consider is how to find the
characteristics of the function $\Psi(0)$ from the measurements of
the mean relative number
$\langle\Psi)|\hat{n}_1-\hat{n}_2|\Psi\rangle$. Let us specify the
case of an odd total number of the particles  (the case of an even
$N$ is discussed below). At $N=2M+1$  and $E_j=0$ the ground state
of the Hamiltonian (\ref{1}) is double-degenerate. The minimum of
the energy equal to $K/8$ is reached for the orthogonal states
$|g_1\rangle=|M+1,M\rangle$ and $|g_2\rangle=|M, M+1\rangle$ as
well as for an arbitrary superposition of these states
$|g\rangle=a|g_1\rangle+b|g_2\rangle$ ($|a|^2+|b|^2=1$). The state
at $t=0$ can be parameterized as
$|\Psi(0)\rangle=\cos(\theta/2)|M+1,M\rangle+\sin(\theta/2)e^{i\varphi}|M,M+1\rangle$
At $\theta\ne 0 ,\pi$ this is the entangled state. The angle
$\theta$ is connected with the initial relative number by the
relation
\begin{equation}\label{3}
\Delta
n(0)\equiv\langle\Psi(0)|\hat{n}_1-\hat{n}_2|\Psi(0)\rangle=\cos
\theta.
\end{equation}
Since this value does not depend on $\varphi$ the phase cannot be
extracted from the result of measurements of the initial relative
number. But the phase $\varphi$ is also an essential
characteristics of the macroscopic state of the Bose condensate.
In particular, the interference pattern emerging under an
overlapping of two BC with the phases $\varphi_1$ and $\varphi_2$
is determined by its relative phase $\phi=\varphi_1-\varphi_2$. We
will show that the value of $\varphi$ can be determined from the
measurements of the final relative number $\Delta n_f\equiv
\langle\Psi(t_f)|\hat{n}_1-\hat{n}_2|\Psi(t_f)\rangle$. To do this
the amplitude of the Josephson coupling should be taken such a
small that the strong inequality $N E_J^{max}/K\ll 1$ be
satisfied. Then at $0<t<t_f$ the system remains in the Fock
regime. In this regime the dynamics of the system is realized
mainly on the states for which
$|\langle\Psi(t)|\hat{n}_1-\hat{n}_2|\Psi(t)\rangle|\leq 1$.
Therefore to find the evolution of the function $\Psi$  one can
use the basis ($|g_1\rangle$, $|g_2\rangle$). Note that the reqime
considered is the same as required for a realization of the Bose
qubit \cite{7}.

It is more convenient to use the unitary transformed basis of
symmetric $|s\rangle=(|g_1\rangle+|g_2\rangle)/\sqrt{2}$ and
antisymmetric $|a\rangle=(|g_1\rangle-|g_2\rangle)/\sqrt{2}$
states. In this basis the wave function of the BC reads as $
 \Psi(t)=s(t)|s\rangle+a(t)|a\rangle
$ Using the   nonstationary Schr\"{o}dinger equation
$i\hbar\dot{\Psi}=H\Psi$ one finds that the functions $s(t)$ and
$a(t)$ satisfy the equations
\begin{equation}\label{5}
  i\hbar\dot{s}=\frac{K}{8}s-\frac{E_j(t)(M+1)}{2} s,
\end{equation}
\begin{equation}\label{6}
  i\hbar\dot{a}=\frac{K}{8}a+\frac{E_j(t)(M+1)}{2} a .
\end{equation}
Integrating equations (\ref{5}), (\ref{6}) we find
 the mean value of the
relative number at the time $t_f$
\begin{eqnarray}\label{12}
  \Delta
  n(t_f)=s^*(t_f)a(t_f)+c.c.=s^*(0)a(0)E^{-2i\Omega}+c.c.=\cr=\cos
  \theta\cos(2\Omega)-\sin\theta\sin(2\Omega)\sin\varphi ,
\end{eqnarray}
where $\Omega=(1/2\hbar)(M+1)\int_0^{t_f} E_J(t') d t'$.

Eq. (\ref{12}) determines the relation between the measured
quantity $\Delta n(t_f)$ and the parameter $\varphi$. One can see
that for entangled initial states the relative number in a final
state depends on the phase $\varphi$ and this phase can be found
from the measurement of $ \Delta  n(t_f)$.

Thus, if one has a system in a reproducible (but unknown) initial
state $|\Psi(0)\rangle$ the parameters $\theta$ and $\varphi$ that
describe this state can be found  from two sets of measurements of
the relative number at $t=0$ and $t\geq t_f$ (under assumption
that the total number of particles $N$ in the condensate is
known). If the total number of the particles is unknown an
additional set of measurements is required: the measurement of the
final relative number occupation with another value of $t_f$.
Using the results of three sets of measurements one can determine
the initial state and find the total number of the particles in
the condensate.

It is necessary to point out  an essential restriction for the
maximum value of $t_f-t_i$. In deriving (\ref{12}) we do not take
into account that the coupling between two condensates induces
small (of order of $E_J(M+1)/K$) but nonzero occupation of the
excited states $|e_k^1\rangle=|M+k+1,M-k\rangle$
$|e_k^2\rangle=|M-k,M+k+1\rangle$ (with $k>0$). Due to such
processes the phases of $s(t)$ and $a(t)$ are shifted from the
values given by the solution of Eqs. (\ref{5}), (\ref{6}). If such
a shift is of order of unity the relation (\ref{12}) is not valid
any more. Nevertheless, one can show that for $t_f-t_i\ll\hbar
K/[E_J^{max}(M+1)]^2$ the phase shifts are very small and Eq.
(\ref{12}) is applicable. The fulfillment of the mentioned
restriction on the value of $t_f-t_i$ is required for the use of
the method of the diagnostics proposed.

Let us now discuss the case of the BC with an even number of
atoms. In the symmetric double-well trap the ground state of the
condensate with $N=2M$ is $|g\rangle=|M,M\rangle$. This state is
nondegenerate and $\Delta n(0)=\Delta n(t_f)=0$. If initially the
system in an excited state
$|e\rangle=a|M+1,M-1\rangle+b|M-1,M+1\rangle$ then $\Delta
n(0)=2(|a|^2-|b|^2)$ can be nonzero, but for $t_f-t_i\ll\hbar
K/[E_J^{max}(M+1)]^2$ the difference $\Delta n(t_f)-\Delta n(0)$
is of order of $M E_J^{max}/K\ll 1$ Such   behavior are in
difference with the case of odd $N$ when the change of $\Delta n$
can be order of unity. This feature can be used for determining
the parity of the number of atoms in BC. We point out again that
this conclusion is for the symmetric relative two BC confining
potential.

If the confining potential is an asymmetric one the Hamiltonian
(\ref{1}) is modified to
\begin{eqnarray}\label{13}
H_1=H_0^a+H_1(t)\cr\equiv\frac{K}{8}(\hat{n}_1-\hat{n}_2)^2-\Delta\mu(\hat{n}_1-\hat{n}_2)
-\frac{E_J(t)}{2}(a_1^+a_2
  +h.c).
\end{eqnarray}
One can see that if $\Delta\mu=K/4$, the ground state of the
system with an even number of the atoms $N=2M$ is
double-degenerate at $E_J=0$, and its wave function can be
presented in the form $|g\rangle=a|M+1,M-1\rangle+b|M, M\rangle$.
This situation is analogous to the symmetric case with odd $N$.
The only difference is that the values of $\Delta n(0)$ and
$\Delta n(t_f)$ given above are counted from $\Delta n=1$. Thus,
under assumption that one can control the value of $\Delta \mu$
with the accuracy $|\Delta\mu-K/4|\ll M E_J^{max}$ the method of
diagnostic of the ground state wave function and the total number
of the atoms suggested is applicable for the BC with even $N$.

Hitherto we have discussed the case of the external Josephson
effect. The case of the internal Josephson effect is also
described by the equation (\ref{13}) (in the rotating frame of
reference) \cite{4}. In this case the expression for the chemical
potential $\Delta\mu$ reads as
\begin{equation}\label{7n}
\Delta\mu=-\delta+\frac{4N\pi
\hbar^2}{m}\tilde{\eta}(a_{11}-a_{22}),
\end{equation}
where $\delta$ is the detuning of the laser field from the
resonant frequency, $a_{11}$ and $a_{22}$, the s-wave scattering
amplitudes of macroscopically  occupied internal states
$|1\rangle$ and $|2\rangle$, correspondingly. In a situation,
where the value of the detuning can be varied  smoothly, one can
achieve the regime of the degenerate ground state of the
Hamiltonian $H_0^a$ both for the even ($\delta=\delta_e$) and for
the odd ($\delta=\delta_o$) number of atoms $N$. In such a regime
one can apply the method of diagnostics of the initial state of BC
proposed in this paper. We note that for the case of the internal
Josephson effect the value of $\overline{\Delta n(t_f)}$  is just
proportional to the expectation value of $M_z(t_f)$ - the
projection of the magnetic momentum of the BC on the axis chosen.
Therefore,  using Eq.(\ref{12}) one can determine the phase
$\varphi$ and the total number of the atoms in the condensate from
the measurement of $M_z(t_f)$.

I would like to acknowledge R.I.Shekhter, S.I.Shevchenko  and
D.V.Fil for the discussion of the results presented in this
article. This work is supported by the INTAS grant No 01-2344.

\end{document}